\documentclass[reprint,a4paper,showpacs,showkeys]{revtex4-1}
\usepackage[utf8]{inputenc}
\usepackage[english]{babel}
\usepackage{graphicx}
\usepackage{amsmath}
\usepackage{amssymb}
\usepackage{textcomp}
\usepackage{subfigure}
\newcommand{\Biz}{$\mathrm{Bi}^{0}$}

\newcommand{\Bipi}{$\mathrm{Bi}^{+}$}
\newcommand{\Bipii}{$\mathrm{Bi}^{2+}$}
\newcommand{\Bipiii}{$\mathrm{Bi}^{3+}$}
\newcommand{\Bipv}{$\mathrm{Bi}^{5+}$}
\newcommand{\Biii}{$\mathrm{Bi}_2$}
\newcommand{\Biiip}{$\mathrm{Bi}_2^{+}$}
\newcommand{\Biiim}{$\mathrm{Bi}_2^{-}$}
\newcommand{\Biiimm}{$\mathrm{Bi}_2^{2-}$}
\newcommand{\BiO}{$\mathrm{BiO}$}
\newcommand{\Biiv}{$\mathrm{Bi}_4^0$}
\newcommand{\Bivpiii}{$\mathrm{Bi}_5^{3+}$}
\newcommand{\SiOiv}{\mbox{$\mathrm{SiO}_4$}}

\newcommand{\BiAlCl}{\mbox{$\mathrm{Bi}_5\!\left(\mathrm{AlCl}_4\right)_3$}}
\newcommand{\atconf}[4]{\mbox{$#1\mathrm{#2}^{#3}_{#4}$}}
\newcommand{\Symm}[4]{\mbox{${}^{#4}\textrm{#1}_{#2}^{#3}$}}
\newcommand{\gamess}{\mbox{GAMESS~(US)}}
\newcommand{\octopus}{\mbox{Octopus}}
\newcommand{\TDDFT}{\mbox{TDDFT}}
\newcommand{\cminv}{$\textrm{cm}^{-1}$}
\begin{document}
\title{%
Possible centers of broadband near-IR luminescence \\
in bismuth-doped solids: \\
$\mathbf{Bi^{+}, Bi_5^{3+}, \text{and~} Bi_4}$
}
\author{V.O.Sokolov}
\email[E-mail:~~]{sokolov@fo.gpi.ac.ru}
\author{V.G.Plotnichenko}
\email[E-mail:~~]{victor@fo.gpi.ac.ru}
\author{E.M.Dianov}
\email[E-mail:~~]{dianov@fo.gpi.ac.ru}
\affiliation{Fiber~Optics~Research~Center of the~Russian~Academy~of~Sciences \\
38~Vavilov~Street, Moscow 119333, Russia}

\begin{abstract}
Subvalent bismuth centers (interstitial \Bipi{} ion, \Bivpiii{} cluster ion,
and \Biiv{} cluster) are examined as possible centers of broadband near-IR
luminescence in bismuth-doped solids on the grounds of quantum-chemical
modeling and experimental data.
\end{abstract}
\maketitle

\section{INTRODUCTION}
Near-IR broadband (1100 -- 1400~nm) luminescence in bismuth-doped glasses
discovered in Ref.~\cite{Murata99} is being studied intensively. By now the
luminescence has been observed in many bismuth-doped glasses, such as
alumosilicate (e.g. \cite{Murata99, Fujimoto01, Dianov06, Khonthon07, Ren06,
Denker09}), alumogermanate (e.g. \cite{Dianov06, Meng05c, Peng05c, Xia06}),
alumoborate (e.g. \cite{Peng05c, Meng05b}), alumophosphosilicate,
alumophosphate, alumophosphoborate (e.g. \cite{Dianov06, Meng05a,
Razdobreev08}), chalcogenide \cite{Yang07, Hughes09}, in several bismuth-doped
crystals (RbPb$_2$Cl$_5$ \cite{Butvina08},
$2\mathrm{MgO}\cdot 2\mathrm{Al}_2\mathrm{O}_3\cdot 5\mathrm{SiO}_2$ cordierite
\cite{Sokolov09}, BaB$_2$O$_4$ \cite{Su09}, BaF$_2$ \cite{Ruan09},
Ba$_2$P$_2$O$_7$ \cite{Peng10}, \BiAlCl{} \cite{Sun11b}), and in FAU-type (e.g.
\cite{Sun09b}) and Y-type \cite{Sun10} zeolites. The bismuth-related IR
luminescence is used successfully in laser amplification and generation (e.g.
\cite{Dianov05, Dianov09}).

However there is no commonly accepted model of the IR luminescence center.
Several models are suggested, such as electronic transitions in \Biz{}
interstitial atoms \cite{Peng09}, in \Bipi{} \cite{Meng05b, Meng05a,
Razdobreev08, Yang07, Sun09b, Sun10, Romanov11}, \Bipii{} \cite{Yang07} and
\Bipv{} \cite{Fujimoto01, Dianov06, Xia06} interstitial ions, in \BiO{}
interstitial molecules \cite{Ren06}, \Biii, \Biiim{} and \Biiimm{} interstitial
dimers \cite{Khonthon07, Meng05c, Sokolov08, Sokolov09, Hughes09}, in other
neutral bismuth clusters \cite{Meng05c}, in \Bivpiii{} cluster ions \cite{Sun10,
Romanov11, Sun11a, Sun11b}, \mbox{$\mathrm{BiO}_4$} complexes with tetrahedral
coordination of the central bismuth ion \cite{Kustov09}, in oxygen-coordinated
complexes formed by pairs of threefold coordinated bismuth atoms
\cite{Denker10}, in complexes formed by \Bipi{} or \Bipii{} substitutional or
interstitial ions with certain defect centers, such as anion vacancies
\cite{Dianov10}.

Recently subvalent bismuth centers such as \Bipi{} single-charged bismuth ion
and \Bivpiii{} cluster ion have attracted considerable attention as possible
sources of the near-IR luminescence \cite{Romanov11, Sun11a, Sun11b}. In this
paper we examine three subvalent bismuth centers basing on experimental data
available and our quantum-chemical modeling. Namely, the above-mentioned
(interstitial) \Bipi{} ion and \Bivpiii{} cluster ion, and (as a novel
suggestion) \Biiv{} cluster are studied.

\section{CALCULATIONS}
Configuration and electronic states of \Bivpiii{} and \Biiv{} clusters were
calculated using \gamess{} quantum-chemical code \cite{gamess}. Firstly, ground
state DFT calculations of the clusters were performed using \mbox{B3LYP1}
functional (Becke+Slater+Hartree-Fock exchange and
Lee-Yang-Parr+Vosko-Wilk-Nusair~(5) correlation --- see \gamess{} manuals
\cite{gamess} for details) and effective core potentials (ECP) and bases of
three types developed in Refs.~\cite{CRENBL, LANL2DZ, SBKJC} with one Huzinaga's
polarization d-type function \cite{GreenBook} added in each basis.

The calculated molecular orbitals were then used to obtain the final molecular
wave functions by means of perturbation theory with multiconfigurational
self-consistent-field reference functions. These final wave functions and
corresponding eigenvalues were used to calculate eigenfunctions and energies of
the ground and exited states of the clusters by method of configuration
interaction in active orbitals space with spin-orbit interaction taken into
account.

Multiconfiguration complete active space self-consistent field (MC-CASSCF) and
multireference configuration interaction (MRCI) were used to calculate
electronic states of \Bivpiii{} and \Biiv{} clusters. Separate CASSCF
calculations were performed for the ground state and the lowest excited state of
each cluster. Then MRCI calculations were made with all CASSCF configurations
with coefficients large enough taken into account (see \gamess{} manuals for
details). Additionally, single and double excitations from these configurations
were allowed.

Evaluative calculation of optical spectra in external electric field and of
ground-state optical absorption spectra (to verify independently the results of
configuration interaction calculations) were performed by time-dependent density
functional theory (\TDDFT) method using \octopus{} program \cite{octopus} and
Hartwigsen-Goedecker-Hutter pseudopotentials \cite{Hartwigsen98} with spin
polarization and spin-orbit interaction taken into account. PBE density
functional \cite{PBE} was used in the ground-state calculation. In the
\octopus{} code, to obtain the linear optical absorption spectrum of the system,
all frequencies of the system are excited by giving certain (small enough)
momentum to the electrons. Then the time-dependent Kohn-Sham equations evolves
in real space for a certain real time \cite{Bertsch00} and the dipole-strength
function (or the photo-absorption cross section) is obtained by a Fourier
transform of the time-dependent dipole moment. Adiabatic LDA approximation is
used in these calculations to describe exchange-correlation effects. The
\octopus{} code uses real-space uniform grid inside the sum of spheres around
each atom of the system (a single one in our case). The sphere radius and the
grid spacing were taken to be was 8.0 and 0.25~\AA, respectively, in our
calculations. The real-time propagation was performed with $2\cdot 10^4$ time
steps with the total simulation time of about 20~fs. The Fourier transform was
performed using third-order polynomial damping (see \cite{octopus} for details
of the code).

\section{ELECTRONIC STATES AND OPTICAL SPECTRA OF FREE BISMUTH SPECIES}
\subsection{$\mathbf{Bi^+}$ bismuth ion}
\begin{figure}
\includegraphics[scale=0.40,bb=-5 -10 605 1010]{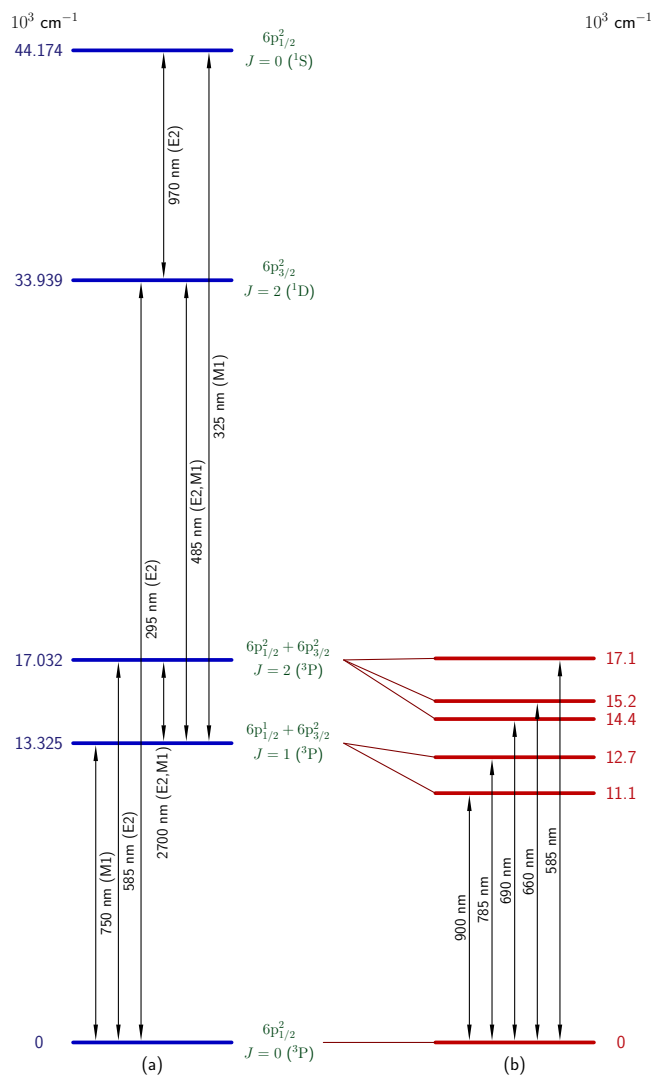}
\caption{%
Experimental states and transitions in (a)~free \Bipi{} ion \cite{Dolk02} and
(b)~\Bipi{} ion in AlCl$_3$--NaCl eutectic \cite{Davis67}.
}
\label{fig:Bi^+_levels}
\end{figure}

The optical absorption and emission of free singly charged positive bismuth
ion,  \Bipi, are well known (see, for example, \cite{Dolk02}). The ground state
and two low-lying excited states with energies about 13325 and 17032~\cminv{}
are known to be formed owing to spin-orbit splitting from the ${}^3\mathrm{P}$
term. The electronic configurations of these three states are
\atconf{6}{p}{2}{1/2} ($J = 0$), \atconf{6}{p}{1}{1/2}+\atconf{6}{p}{2}{3/2} ($J
= 1$), and \atconf{6}{p}{1}{1/2}+\atconf{6}{p}{2}{3/2} ($J = 2$), respectively.
The subsequent excited states with much greater energy (about 33940 and
44175~\cminv) correspond to ${}^1\mathrm{D}$ and ${}^1\mathrm{S}$ terms with
electronic configurations \atconf{6}{p}{2}{3/2} ($J = 2$) and
\atconf{6}{p}{2}{3/2} ($J = 0$), respectively. Obviously, all the electric
dipole (E1) transitions between the states just listed are forbidden, only
magnetic dipole (M1) and electric quadrupole (E2) transitions being allowed. In
particular, the ground state of \Bipi{} ion is characterized by weak absorption
at wavelengths about 750, 585 and 295~nm corresponding to M1 and E2 transitions
from the ground state to three lowest excited states. The most long-wavelength
luminescence in \Bipi{} ion corresponds to the M1 transition from the first
excited state to the ground one (Fig.~\ref{fig:Bi^+_levels}~(a)). Obviously,
there is no Stokes shift in free ion.

In Ref.~\cite{Davis67} absorption spectra of \Bipi{} ion in chlorides melts
have been studied and interpreted basing on crystal field theory. According to
this study, two lowest excited states of the ion are split by the crystal
field in two and three levels, respectively. As a result, the transition energy
decreases significantly, and bands around 900, 785, 660--690, and 585~nm are
observed in the absorption spectrum (Fig.~\ref{fig:Bi^+_levels}~(b)).

Unfortunately, calculation of \Bipi{} ion spectra (both free ion and one in
electrostatic field) with the spin-orbit interaction taken into account presents
severe computational problems using \gamess{} program. Our conservative estimates 
and evaluative calculations performed by the \TDDFT{} method using \octopus{} 
code causes us to anticipate that the wavelength of the lowest-energy transition 
does not exceed 1000~nm essentially. Such a conclusion is consistent with the 
results of Ref.~\cite{Davis67}.
\begin{figure*}
\subfigure[]{%
\includegraphics[scale=0.435,bb=150 0 690 800]{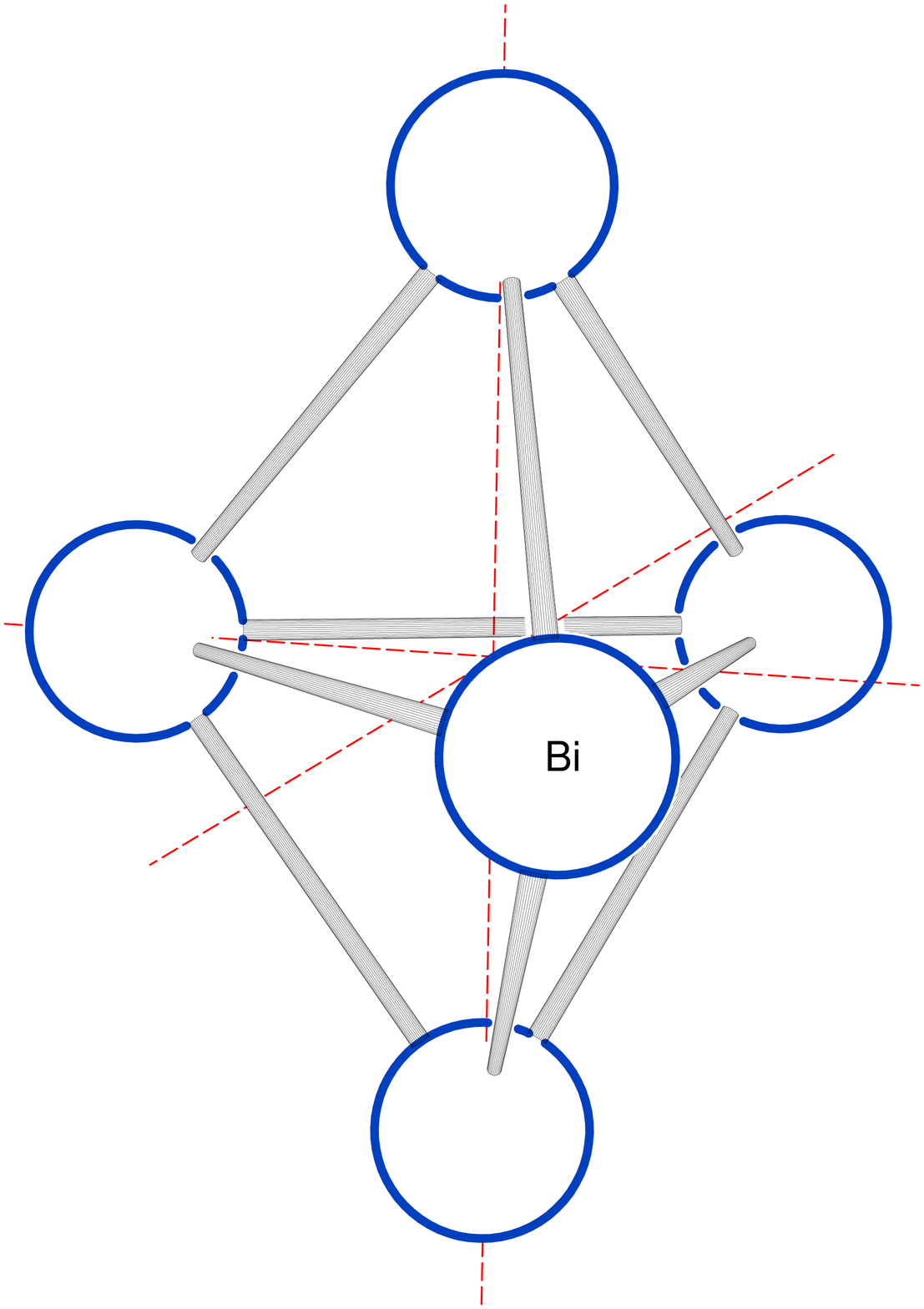}
\label{fig:Bi_5^3+-a}
}
\qquad\quad
\subfigure[][]{%
\includegraphics[scale=0.435,bb=150 0 690 800]{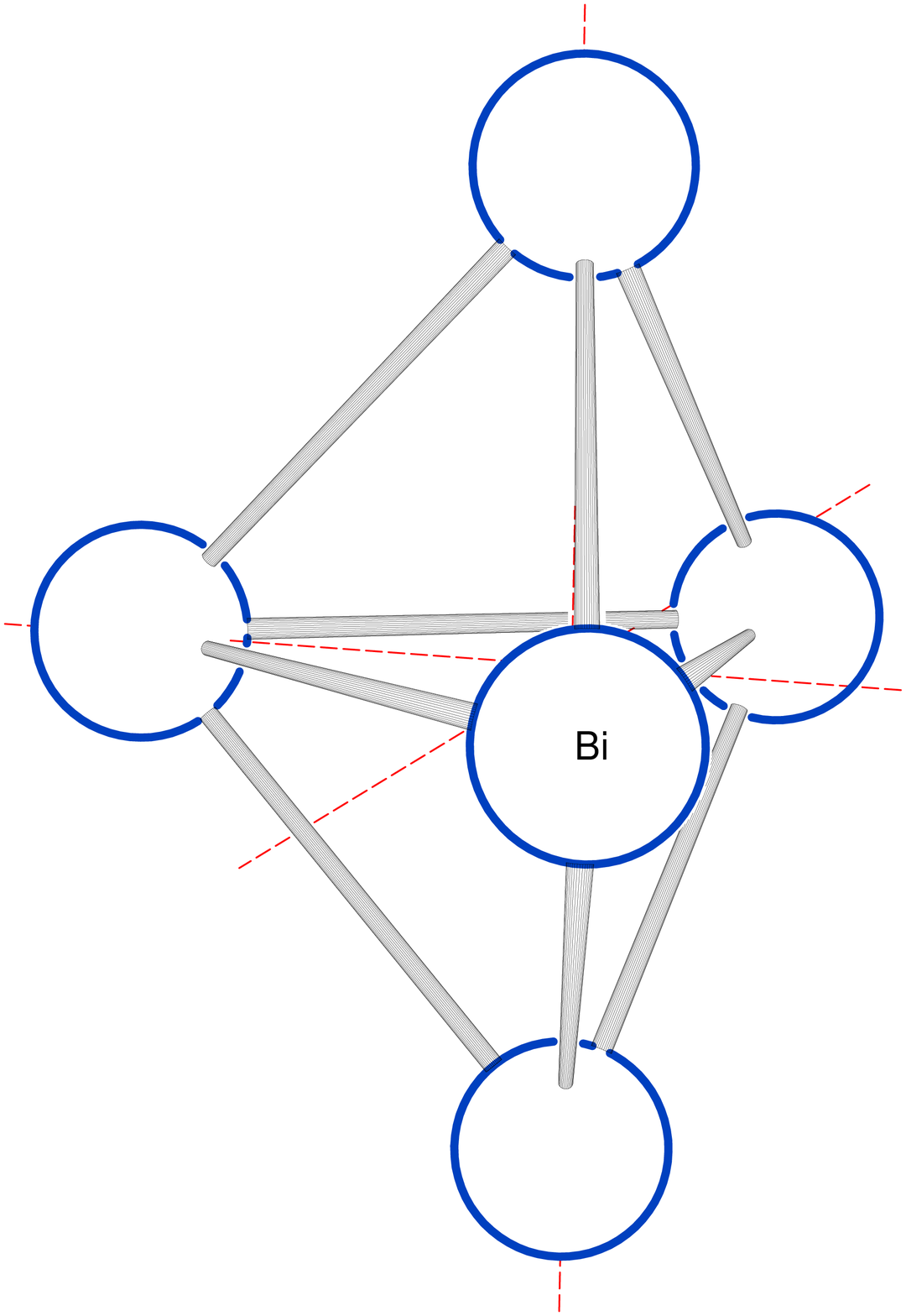}
\label{fig:Bi_5^3+-b}
}
\caption{%
Calculated configurations of free \Bivpiii{} cluster ion:
\subref{fig:Bi_5^3+-a}~ground-state singlet (\Symm{D}{3h}{}{}{} symmetry),
\subref{fig:Bi_5^3+-b}~first excited-state triplet (\Symm{C}{2v}{}{}{}
symmetry).
}
\label{fig:Bi_5^3+}
\end{figure*}

\subsection{$\mathbf{Bi_5^{3+}}$ cluster ion}
\begin{figure}
\includegraphics[scale=0.35,bb=0 -10 710 1010]{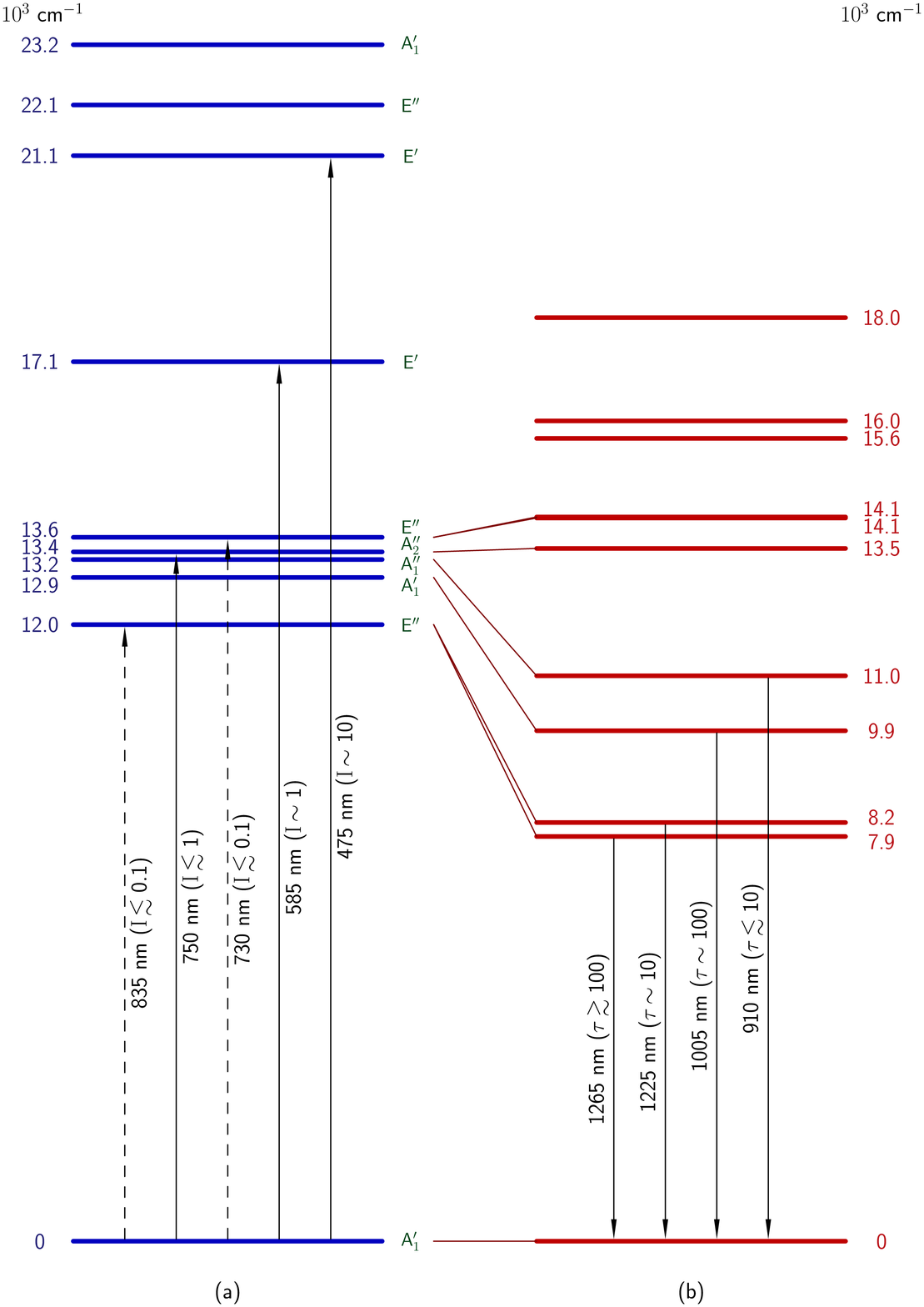}
\caption{%
Calculated states, excitation transition rates and luminescence lifetimes in
free \Bivpiii{} cluster ion: (a)~ground state, (b)~1-st excited state
(transition rates and lifetimes in relative units).
}
\label{fig:Bi_5^3+_levels}
\end{figure}

\begin{figure}
\includegraphics[scale=0.525,bb=70 285 545 815]{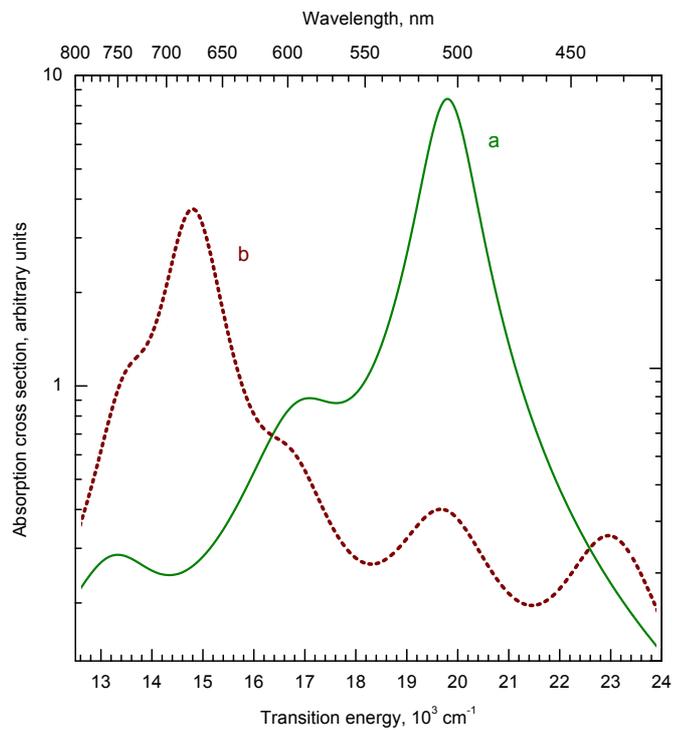}
\caption{%
Absorption cross section of (a)~\Bivpiii{} and (b)~\Biiv{}
clusters calculated by \TDDFT{} method.
}
\label{fig:By_5^3+_Bi_4_TDDFT}
\end{figure}

Our calculations prove the electronic ground state of \Bivpiii{} cluster ion to
be a singlet. In this state, the ion symmetry is found to be \Symm{D}{3h}{}{}{}
(Fig.~\ref{fig:Bi_5^3+}~(\subref{fig:Bi_5^3+-a})). Interatomic distances
Bi$_{ax}\!\relbar$Bi$_{eq}$ and Bi$_{eq}\!\relbar$Bi$_{eq}$ between axial,
Bi$_{ax}$, and equatorial, Bi$_{eq}$, bismuth atoms are 3.089~\AA{} and
3.325~\AA{}, respectively. This agrees well with both experimental and
calculated data for free \Bivpiii{} cluster ion \cite{Day00}. In
Fig.~\ref{fig:Bi_5^3+_levels}~(a) are shown the calculated scheme of the
\Bivpiii{} cluster ion levels and transitions between them. In the ground state
configuration, the electronic states are classified according to irreducible
representations of the \Symm{D}{3h}{}{}{} group. From the ground state with the
\Symm{A}{1}{\prime}{} symmetry E1 transitions are allowed to excited states with
\Symm{E}{}{\prime}{}{} and \Symm{A}{2}{\prime\prime}{} symmetry, giving rise to
absorption at 750, 585 and 450~nm wavelengths (shown by solid arrows in the
Fig.~\ref{fig:Bi_5^3+_levels}). When perturbed by an external electrostatic
(crystalline) field, E1 transitions from the ground state to the excited states
with \Symm{E}{}{\prime\prime}{}{} symmetry become slightly allowed corresponding
to absorption at wavelengths of about 835 and 730~nm (displayed as dashed arrows
in Fig.~\ref{fig:Bi_5^3+_levels}). It should be noted that in the absence of
spin-orbit interaction all the excited states in Fig.~\ref{fig:Bi_5^3+_levels}
correspond to the first excited triplet. These results are in good agreement
both with the experimental data available on \Bivpiii{} cluster ions optical
absorption \cite{Day00, Bjerrum67}, and with the results of our \TDDFT{}
calculations using the \octopus{} code shown in
Fig.~\ref{fig:By_5^3+_Bi_4_TDDFT}. As well those agree reasonably with
experimental data for \BiAlCl{} crystal \cite{Sun11b}.

Fig.~\ref{fig:Bi_5^3+}~(\subref{fig:Bi_5^3+-b}) shows the calculated
configuration of \Bivpiii{} cluster ion in the lowest excited state. To our
knowledge, the calculation of configuration of this ion in an excited state is
performed for the first time. Being excited to this state, \Bivpiii{} cluster
ion undergoes deformation corresponding to \Symm{E}{}{\prime\prime}{}{}
irreducible representation of the \Symm{D}{3h}{}{}{} group. As a result, the
cluster ion symmetry is reduced to \Symm{C}{2v}{}{}. One of the
Bi$_{ax}\!\relbar$Bi$_{eq}$ interatomic distances is increased to 3.363~\AA, the
other two distances are decreased to 3.043~\AA, and corresponding
Bi$_{eq}\!\relbar$Bi$_{eq}$  distances become 3.332~\AA{} and 3.362~\AA{},
respectively. Calculated levels of \Bivpiii{} cluster ion in the configuration
of its lowest excited state, transitions to the ground state and relative
lifetimes are shown in Fig.~\ref{fig:Bi_5^3+_levels}~(b). In our calculations
two E1 transitions to the ground state are found in this configuration. Those
lead to two bands of IR luminescence in 1200 -- 1290~nm range with lifetimes
differing by an order of magnitude, which may be called ''slow'' and
''fast'' components (unfortunately, only relative lifetimes may be estimated in
our calculations). It should be remarked that such two components of
bismuth-related IR luminescence have been observed repeatedly (see e.g.
Refs.~\cite{Dianov06, Denker09, Sun11a}).

Furthermore, there are luminescence bands near 1000 and
900~nm (as well ''slow'' and ''fast'', respectively). These results are also
confirmed in our \TDDFT{} calculations.
\begin{figure*}
\subfigure[]{%
\includegraphics[scale=0.311,bb=0 -50 670 730]{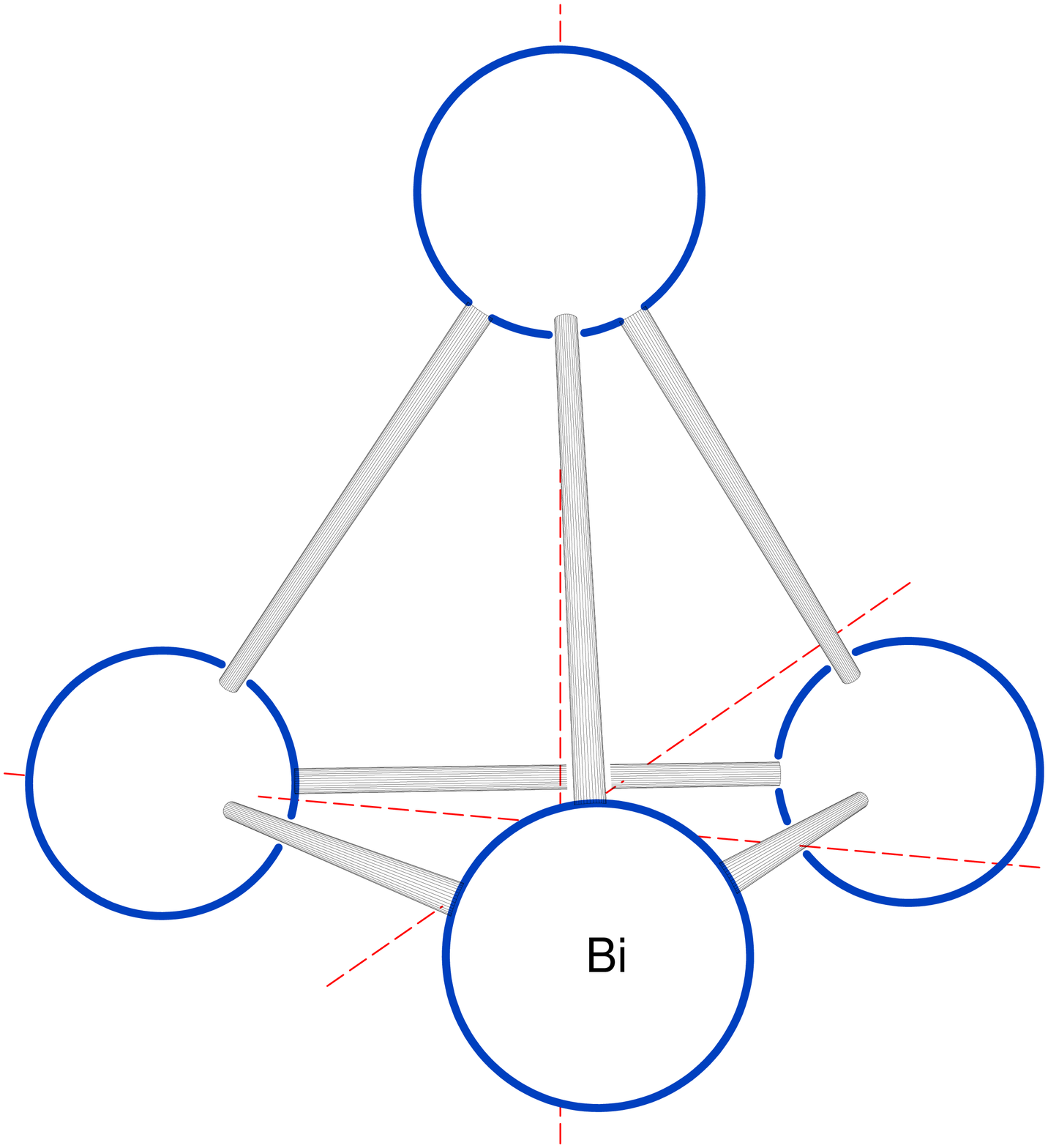}
\label{fig:Bi_4-a}
}
\qquad\qquad
\subfigure[]{%
\includegraphics[scale=0.349,bb=0 -50 590 650]{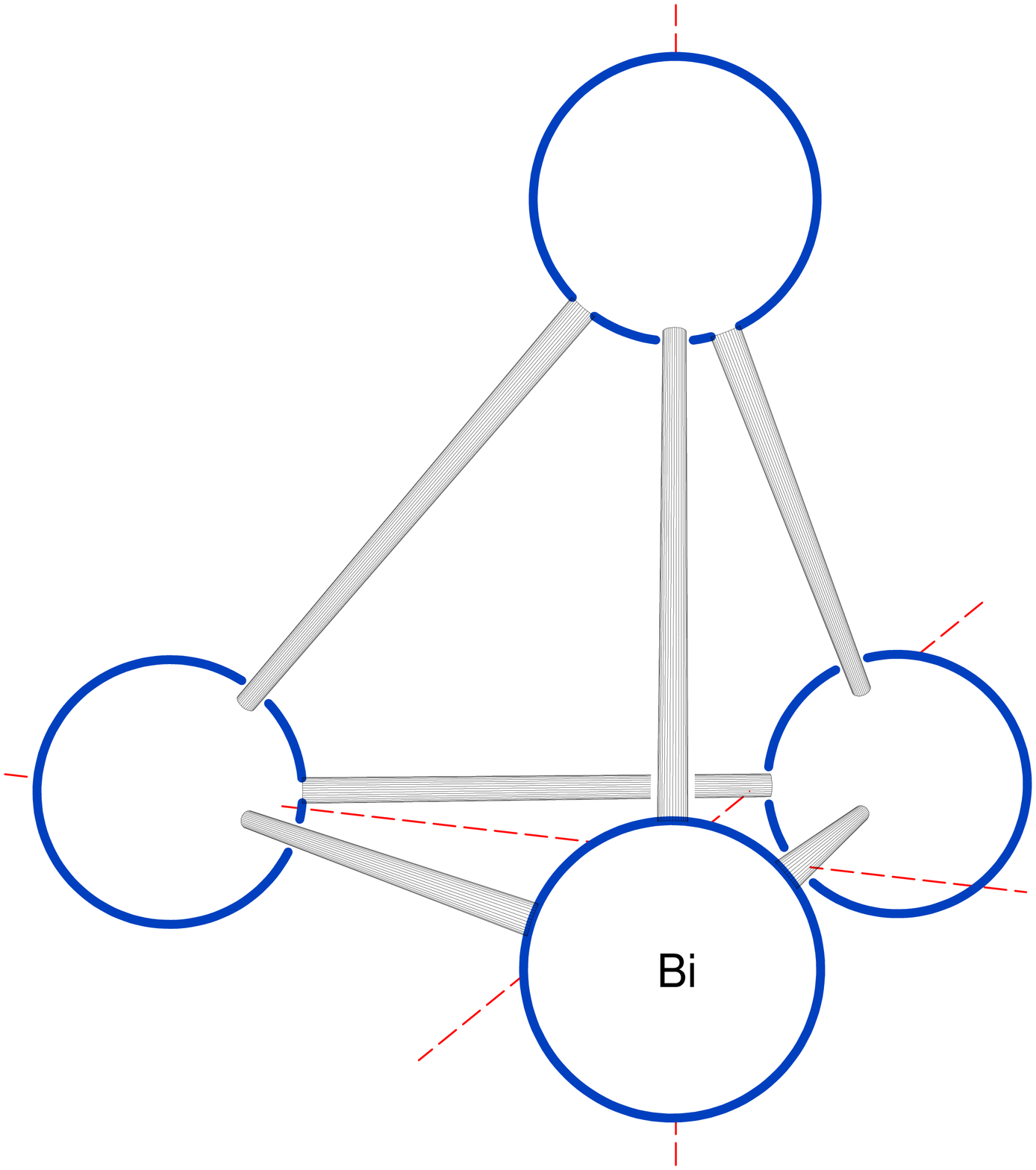}
\label{fig:Bi_4-b}
}
\caption{%
Calculated configuration of free \Biiv{} cluster:
\subref{fig:Bi_4-a}~ground-state singlet (\Symm{T}{d}{}{}{} symmetry),
\subref{fig:Bi_4-b}~first excited-state triplet (\Symm{C}{2v}{}{}{}
symmetry).
}
\label{fig:Bi_4}
\end{figure*}

\subsection{$\mathbf{Bi_4^0}$ cluster}
\begin{figure}
\includegraphics[scale=0.35,bb=-5 -10 695 1010]{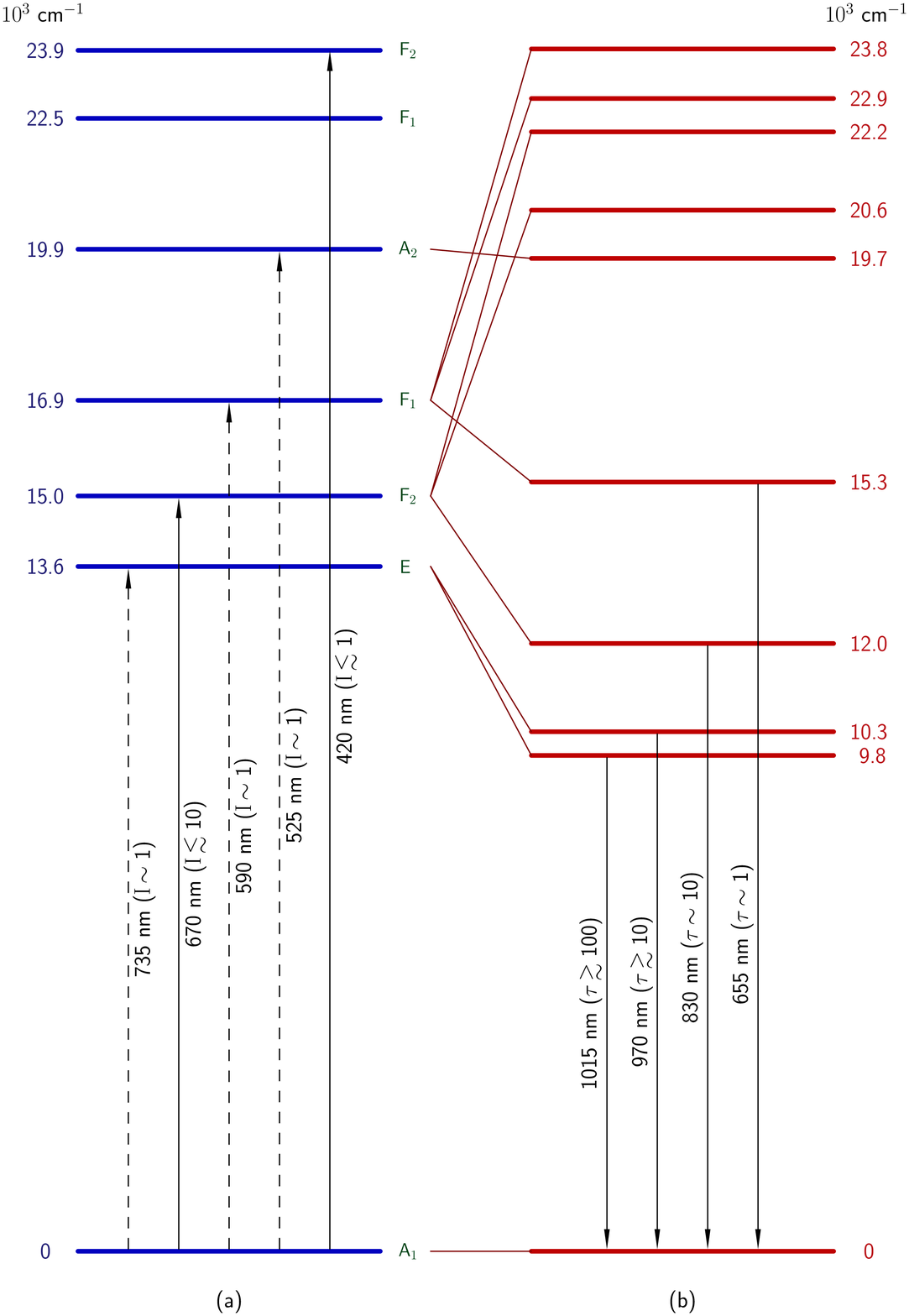}
\caption{%
Calculated states, excitation transition rates and luminescence lifetimes in
free \Biiv{} cluster: (a)~ground state, (b)~1-st excited state (transition
rates and lifetimes in relative units).
}
\label{fig:Bi_4_levels}
\end{figure}

According to our calculation the electronic ground state of \Biiv{} cluster
turns out to be a singlet. As seen from
Fig.~\ref{fig:Bi_4}~(\subref{fig:Bi_4-a}), in this state the \Biiv{} cluster is
a regular tetrahedron (\Symm{T}{d}{}{}{} symmetry).
Bi$\relbar$Bi interatomic distances are found to be 3.104~\AA{} in agreement
with the results of Ref.~\cite{Balasubramanian92}.
Fig.~\ref{fig:Bi_4_levels}~(a) presents the calculated scheme of levels,
transitions between those and the relative intensity of the transitions in the
\Biiv{} cluster. The electronic states of the cluster in the ground state
configuration correspond to \Symm{T}{d}{}{}{} group irreducible representations.
Thus, the ground state has \Symm{A}{1}{}{}{} symmetry. From the ground state
E1 transitions only to the excited state with \Symm{F}{2}{}{}{} symmetry are
allowed. Those correspond to absorption bands near 670 and 420~nm (solid arrows
in Fig.~\ref{fig:Bi_4_levels}). In an external electrostatic field E1
transitions from ground state to the states with \Symm{A}{2}{}{},
\Symm{F}{1}{}{}, and \Symm{E}{}{}{}{} symmetry become slightly allowed as well,
and absorption at wavelengths of about 735, 590 and 525~nm arises (dashed arrows
in Fig.~\ref{fig:Bi_4_levels}). The first four excited states are formed due to
spin-orbit splitting of the \Symm{F}{2}{}{3}{} triplet state, and the next ones
arise from \Symm{F}{2}{}{1}{} and \Symm{F}{1}{}{1}{} singlet states.

The energies of states and transitions in \Biiv{} cluster obtained in our
calculation agree well with the results of the calculations in
Ref.~\cite{Balasubramanian92}. However the results differ in the order of
levels: the states with \Symm{E}{}{}{}{} and  \Symm{A}{2}{}{}{} symmetries turn
out to be interchanged. On the other hand, the transition energies obtained in
quantum-chemical calculations, both our and that of
Ref.~\cite{Balasubramanian92}, agree closely with our \TDDFT{} calculation
(Fig.~\ref{fig:By_5^3+_Bi_4_TDDFT}).

Excitation to the lowest excited state with \Symm{E}{}{}{}{} symmetry is
accompanied by rearrangement of the cluster configuration resulting in the
cluster symmetry reduced to \Symm{C}{2v}{}{}. The length of one of the
tetrahedron edges (Bi$\relbar$Bi distance) is increased to 3.378~\AA{} and the
length of the opposite edge is reduced to 3.005~\AA, while the other
Bi$\relbar$Bi distances are reduced to 3.082~\AA{}
(Fig.~\ref{fig:Bi_4}~(\subref{fig:Bi_4-b})).

In Fig.~\ref{fig:Bi_4_levels}~(b) are shown the calculated levels together with
transitions to the ground state and their relative lifetimes for the lowest
excited state configuration of \Biiv{} cluster. According to the calculations,
in this configuration the most long-wavelength luminescence in the 950 --
1040~nm range contains two components (''slow'' and ''fast'' in the same degree
as above) corresponding to E1 transitions from the first and second excited
states to the ground one. Furthermore, E1 transitions from two higher excited
states to the ground state can lead to the ''fast'' luminescence near 830 and
660~nm. Our \TDDFT{} calculations with the \octopus{} code yield similar
results.
\begin{figure}
\includegraphics[scale=0.325,bb=35 150 805 915]{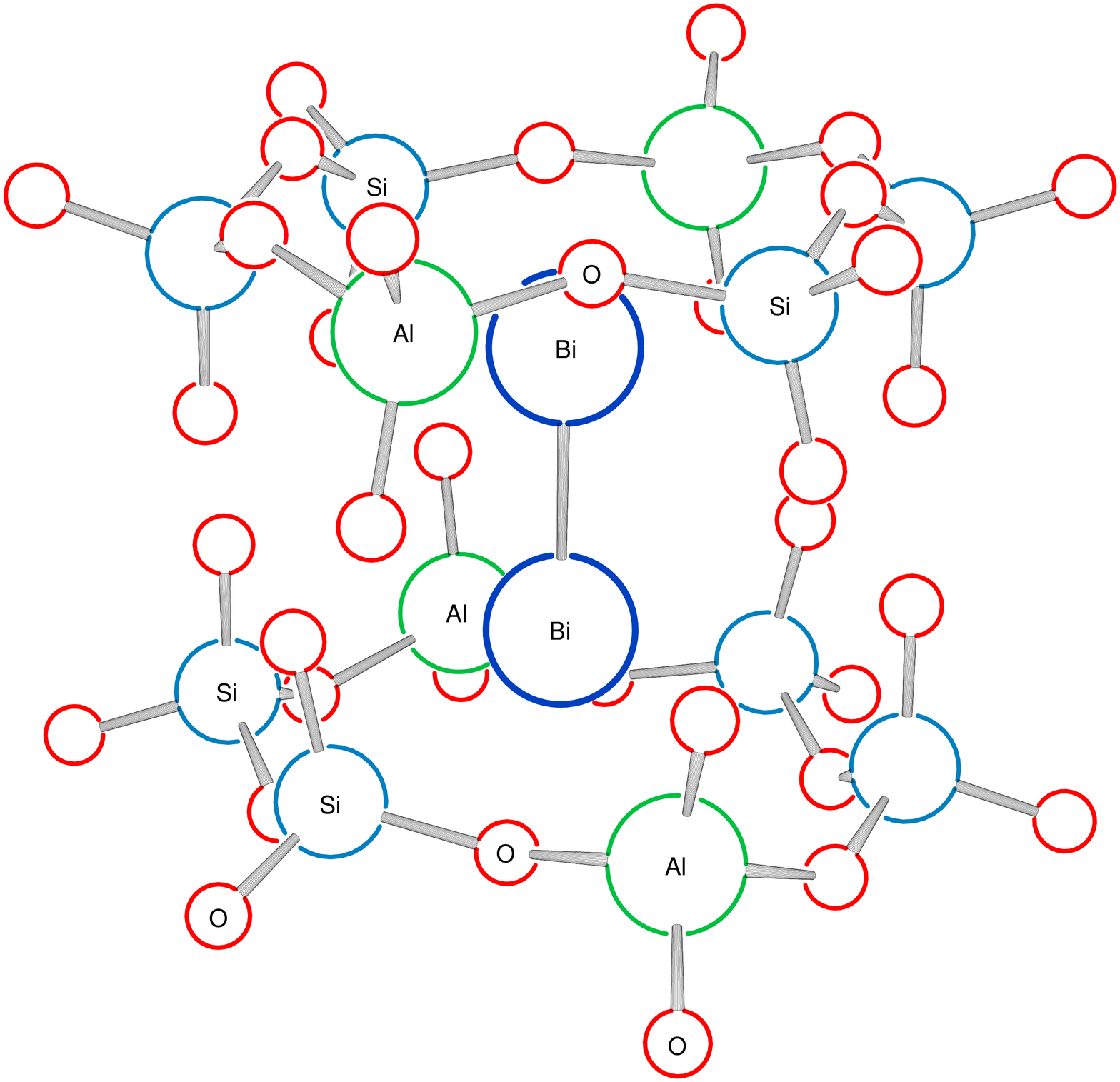}
\caption{%
Calculated configuration of \Biiip{} dimer in six-member ring interstitial of
alumosilicate glass network.
}
\label{fig:Bi_2+_in_rings}
\end{figure}

\section{BISMUTH SPECIES IN SILICATE GLASS NETWORK}
To study the possibilities of occurrence of \Bipi{} ion, \Bivpiii{} cluster
ion, and \Biiv{} cluster in silicate glass network, we performed
quantum-chemical modeling of these centers in ring interstitial sites of silica
glass network using cluster approach. In the clusters there are either one or
two six-member rings formed by four \SiOiv{} tetrahedra. Dangling bonds of the
outer oxygen atoms in the clusters were saturated with hydrogen atoms. The
bismuth species was placed initially at the center of the ring(s) and then the
complete geometry optimization was performed. Details of computational
techniques were described above. Standard \mbox{3-21G} basis was used for
saturating hydrogen atoms.

\subsection{$\mathbf{Bi^+}$ bismuth ion}
In our calculations equilibrium position of \Bipi{} ion in silicate glass
network is found in interstitial sites formed by six-member rings, \Bipi{} ion
being located near the rings axis between the rings planes. However, the
calculations prove this position of \Bipi{} ion to be not stable enough: even
relatively small displacement of the ion (not more than 10~\%{} of the minimal
distance to neighboring atoms) leads to further drift apart from the equilibrium
position. Subsequent process is determined by the environment. Thus, in the
regular ring interstitial \Bipi{} ion moves towards one of the nearest oxygen
atoms in the surrounding rings and threefold coordinated bismuth atom and
threefold coordinated silicon atom are formed. If there are another \Bipi{}
ion(s) or \Biz{} atom(s) nearby, interstitial dimers or larger bismuth clusters
are likely to arise instead. As an example, calculated position of \Biiip{}
dimer in six-member rings interstitial of alumosilicate network is shown in
Fig.~\ref{fig:Bi_2+_in_rings}. Some evidence is found in our calculations that
\Bipi{} ion may be stabilized in interstitial site of the silicate glass network
by neighboring defects, such as oxygen vacancies.

Thus, although single interstitial \Bipi{} ions may occur in the silicate glass
network, their stability is poor and significant concentration of such ions
seems hardly probable. On the other hand, small bismuth clusters or \Bipi{} ---
defect complexes may be formed in the network interstitials, such clusters and
complexes being quite stable.

It is notable that the luminescence Stokes shift which vanishes in free \Bipi{}
ion should be expected to be low in the interstitial ion as well, but this is
not the case for the above-mentioned clusters or complexes.

\subsection{$\mathbf{Bi_5^{3+}}$ cluster ion}
Geometric sizes of \Bivpiii{} cluster ion are obviously too large to allow the
ion to be incorporated in ring interstitial sites of silicate glasses. On the
other hand, owing to significant electric charge the ion is very likely to react
with the neighboring electronegative atoms.

Our calculations performed using the above-described cluster models confirm
these considerations. It is found that \Bivpiii{} cluster ion reacts with the
bridging oxygen atoms of the network to form interstitial \Biiv{} cluster and
single \Bipiii{} ion nearby. The latter is then built into the glass network
forming threefold coordinated bismuth atom. The resulting structure is shown in
Fig.~\ref{fig:Bi_4_in_rings}.

It should be realized that even \Bivpiii{} ion is not able to react with
neighboring network atoms for whatever reason, the $3+$ charge state of the
$\mathrm{Bi}_5$ cluster turns out to be unstable with respect to electron
capture. We have calculated electron affinity of \Bivpiii{} cluster ion by the
above-described quantum-chemical methods in DFT level of theory and obtained
closely agreed values for vertical and adiabatic electron affinities:
$14.2\text{~eV} \lesssim E\!A_{v} \lesssim E\!A_{a} \lesssim 14.6\text{~eV}$.

These values significantly exceed not only the electron affinity of the
silicate network ($\lesssim 1.2 \text{~eV}$), but its band gap width ($\lesssim
9\text{~eV}$) as well. Hence \Bivpiii{} cluster ion should capture electron in
silicate host, the capture being possible not only from the conduction band, but
from the valence band as well.

Thus, interstitial \Bivpiii{} cluster ions occurrence should be recognized to be
impossible in silicate hosts.

However, it seems reasonable to guess that \Bivpiii{} cluster ion can occur in
zeolite hosts due to extremely large size of their cage-like interstitials and
stabilizing action of $\left(\mathrm{AlO}_4\right)^{1-}$ tetrahedra similar to
such action of $\left(\mathrm{AlCl}_4\right)^{1-}$ ions in \BiAlCl{} crystal
\cite{Sun11b}.

\subsection{$\mathbf{Bi_4^0}$ cluster}
\begin{figure}
\includegraphics[scale=0.325,bb=25 0 795 775]{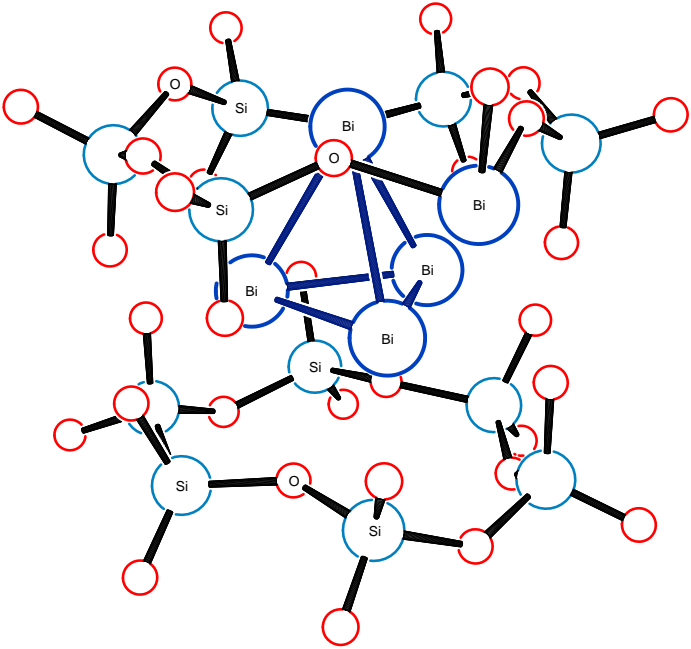}
\caption{%
Calculated configuration of \Biiv{} cluster in six-member ring interstitial of
silica glass network.
}
\label{fig:Bi_4_in_rings}
\end{figure}

The calculations have proved \Biiv{} clusters to be able to occur in ring
interstitials of silicate networks, despite the large size of the cluster. One
such example is above-described formation of \Biiv{} cluster from \Bivpiii{}
cluster ion. However threefold coordinated bismuth atom does not need to be
present in the surrounding rings: according to our modeling, a configuration
quite similar to that shown in Fig.~\ref{fig:Bi_4_in_rings} arises as well in
purely-silicate and alumosilicate rings. Such positions of \Biiv{} clusters in
ring interstitials are found to be extremely stable. Even relatively large
displacement and (or) rotation of the \Biiv{} cluster, accompanied by its atoms
shifted up to 50~\%{} of interatomic distance do not make the above-mentioned
configuration be destroyed.

Thus, \Biiv{} clusters can occur in the ring interstitial of the pure-silicate
network, and probably in certain binary silicate networks as well, such an
interstitial position being extremely stable. Together with the fact the \Biiv{}
cluster binding energy is known to be abnormally high \cite{Yuan08}, this
suggests the possibility of high concentration of such clusters in silicate
hosts.

\section{CONCLUSIONS}
Both the experimental data available and the results of our calculations allow
draw some conclusions concerning the origin of the above-mentioned IR
luminescence centers in the bismuth-containing systems.
\begin{enumerate}
\item For \Bipi{} singly-charged positive ion:
\begin{enumerate}
\item the lowest-energy transition wavelength does not exceed 1000~nm;
\item all the relatively long-wavelength transitions occur between
electronic states originating (due to spin-orbit interaction) from the same
${}^3\mathrm{P}$ term and hence intensities of both absorption and luminescence
transitions are of the same order;
\item Stokes shift is low or even vanishes;
\item although there are equilibrium interstitial positions of \Bipi{} ion
in certain solids, single \Bipi{} seems to be not stable enough, at least in
oxide glasses networks.
\end{enumerate}
Thus, single \Bipi{} ion can be responsible for certain optical absorption bands
in bismuth-containing systems, but can not be considered as a center of the IR
luminescence.
\item For \Bivpiii{} cluster ion:
\begin{enumerate}
\item IR luminescence spectrum accords well enough with the experimentally
observed luminescence in bismuth-containing systems;
\item there are a number of bands in the absorption spectrum corresponding to
the absorption observed in experiment;
\item in the absorption spectrum of free \Bivpiii{} cluster ion there is no band
at 500~nm, the most intense in experimental spectra of bismuth-containing
systems;
\item large size, high electrical charge and high electron affinity of the
\Bivpiii{} cluster ion make it practically impossible for the ion to occur in a
solid host other than, supposedly, zeolite.
\item in silicate glass the cluster ion can react with bridging oxygen atoms of
the glass network forming threefold coordinated bismuth atom bound in the
network and interstitial \Biiv{} cluster.
\end{enumerate}
Thus, \Bivpiii{} cluster ion can be responsible for a number of absorption
and IR luminescence bands in certain bismuth-containing systems, namely, ionic
solutions and, possibly, zeolites, but not for those in silicate glasses.
\item For \Biiv{} cluster:
\begin{enumerate}
\item in free \Biiv{} cluster, IR luminescence occur only near 1000~nm
wavelength, but under the influence of crystal field the wavelength can increase
(up to 1100 -- 1150~nm according to our estimations);
\item in the absorption spectrum of the cluster there are a number of bands
corresponding to the absorption observed in experiment including the band
at 500~nm. However, the band near 670~nm is the most intense, which is hardly
consistent with experimental data;
\item there are equilibrium positions of \Biiv{} cluster in sufficiently large
interstitial sites in solid hosts (e.g. in the ring interstitials of oxide
glasses networks), these positions being quite stable.
\end{enumerate}
Thus, \Biiv{} cluster can be responsible for a number of absorption and IR
luminescence bands in the bismuth-containing systems.
\item If in a bismuth-containing system both \Bipi{} ions and \Bivpiii{} cluster
ions or (and) \Biiv{} clusters occur simultaneously, one would expect mutual
excitation transfer between the centers of different types.
\end{enumerate}

Thus, none of the \Bipi, \Bivpiii, and \Biiv{} centers alone can not give rise
to the experimental IR luminescence and absorption spectra typical for the
bismuth-doped solids. On the other hand, these spectra of ionic solutions are
likely to be caused by \Bipi{} and \Bivpiii{} ions together.

It is our opinion that the experimental data available are consistent with the
following speculations involving the centers under consideration. In solid
hosts, most likely both \Bipi{} ions and \Biiv{} clusters occur in interstitial
sites (the latters in sufficiently large interstitials, in the ring ones
particularly). Moreover, it is conceivable that interstitial \Bipi{} ions could
form pairs with each other or with interstitial \Biz{} atoms. In ionic
solutions, \Bipi{} ions and \Bivpiii{} cluster ions most likely occur.
Excitation transfer from \Bipi{} ions to other bismuth centers is expected to
proceed in all the cases.

On the other hand, it should be emphasized that no experimental data nor
calculation results suggest that in the bismuth-doped systems there are no
IR luminescence centers other than those considered in this paper. In
particular, we persist in the belief that negatively charged bismuth dimers,
\Biiim{} and \Biiimm{} \cite{Sokolov08, Sokolov09}, and dimers (or even more
complex clusters) of \Bipi{} or \Bipii{} ions and some intrinsic defect centers
\cite{Dianov10} are quite possible to occur in bismuth-doped solids. Moreover,
it seems reasonable to expect that bismuth centers of several types may be
formed concurrently in these solids. The assumptions concerning existence of
several types of IR luminescence centers in bismuth-doped solids is supported by
significant variation of both absorption spectra and IR luminescence spectral
and temporal dependencies, with preparation conditions etc. observed in various
systems.

%
%

\end{document}